\documentclass[twocolumn,english]{IEEEtran}
\usepackage[T1]{fontenc}
\usepackage{babel}
\usepackage{array}
\usepackage{textcomp}
\usepackage{multirow}
\usepackage{amsmath}
\usepackage{amssymb}
\usepackage{graphicx}
\usepackage{epstopdf}
\usepackage{cite}
\usepackage[unicode=true,
 bookmarks=true,bookmarksnumbered=true,bookmarksopen=true,bookmarksopenlevel=1,
 breaklinks=false,pdfborder={0 0 0},backref=false,colorlinks=false]
 {hyperref}
\hypersetup{pdftitle={Your Title},
 pdfauthor={Your Name},
 pdfpagelayout=OneColumn, pdfnewwindow=true, pdfstartview=XYZ, plainpages=false}

\makeatletter

\providecommand{\tabularnewline}{\\}

\ifCLASSOPTIONcompsoc
\usepackage[caption=false,font=normalsize,labelfont=sf,textfont=sf]{subfig}
\else
\usepackage[caption=false,font=footnotesize]{subfig}
\fi

\@ifundefined{showcaptionsetup}{}{%
 \PassOptionsToPackage{caption=false}{subfig}}
\usepackage{subfig}
\makeatother

\begin{document}

\title{A Generic Self-Evolving Neuro-Fuzzy Controller based High-performance
Hexacopter Altitude Control System }

\author{Md Meftahul~Ferdaus,~\IEEEmembership{Student Member,~IEEE,} Mahardhika
~Pratama,~\IEEEmembership{Member,$\:$IEEE,} Sreenatha G. ~Anavatti,~and~Matthew
A. Garratt%
\thanks{Md Meftahul~Ferdaus is with~the School of Engineering and Information
Technology, University of New South Wales at the Australian Defence
Force Academy, Canberra, ACT 2612, Australia, e-mail: \protect\href{mailto:m.ferdaus@student.unsw.edu.au}{m.ferdaus@student.unsw.edu.au}.%
}%
\thanks{Mahardhika Pratama~is with the School of Computer Science and Engineering,
Nanyang Technological University, Singapore, 639798, Singapore, e-mail:
\protect\href{mailto:mpratama@ntu.edu.sg}{mpratama@ntu.edu.sg}.%
}%
\thanks{Sreenatha G. ~Anavatti~ is with the School of Engineering and Information
Technology, University of New South Wales at the Australian Defence
Force Academy, Canberra, ACT 2612, Australia, e-mail: \protect\href{mailto:s.anavatti@adfa.edu.au}{s.anavatti@adfa.edu.au}.%
}%
\thanks{Matthew A. Garratt~ is with the School of Engineering and Information
Technology, University of New South Wales at the Australian Defence
Force Academy, Canberra, ACT 2612, Australia, e-mail: \protect\href{mailto:M.Garratt@adfa.edu.au}{M.Garratt@adfa.edu.au}.%
}%
}
\maketitle
\begin{abstract}
Nowadays, the application of fully autonomous system like rotary wing
unmanned air vehicles (UAVs) is increasing sharply. Due to the complex
nonlinear dynamics a huge research interest is witnessed in developing
learning machine based intelligent, self-organizing evolving controller
for these vehicles notably to address the system's dynamic characteristics. In this work, such an evolving controller namely
Generic-controller (G-controller) is proposed to control the altitude
of a rotary wing UAV namely hexacopter. This controller can work with
very minor expert domain knowledge. The evolving architecture of this
controller is based on an advanced incremental learning algorithm
namely Generic Evolving Neuro-Fuzzy Inference System (GENEFIS). The
controller does not require any offline training, since it starts
operating from scratch with an empty set of fuzzy rules, and then
add or delete rules on demand. The adaptation laws for the consequent
parameters are derived from the sliding mode control (SMC) theory.
The Lyapunov theory is used to guarantee the stability of the proposed
controller. In addition, an auxiliary robustifying control term is
implemented to obtain a uniform asymptotic convergence of tracking
error to zero. Finally, the G-controller's performance evaluation
is observed through the altitude tracking of a UAV namely hexacopter
for various trajectories.\end{abstract}

\section{Introduction\label{sec:Introduction}}

Unmanned air vehicles (UAVs) are aircraft with no pilot on board.
UAV's autonomy shifts from partial to complete, which starts from
the human operator based partial remote control to completely self-governing
control by onboard computers. Autonomy empowers UAVs to perform some
tasks very well where human contribution would be hazardous, or too
tedious. In addition, due to their light weight, fuel efficiency and
easier maintenance than their manned counterparts, they are getting
prevalence day by day with huge applicability in both military and
civil sectors \cite{ferdaus2017fuzzy}. In connection to the wing
types, UAVs are normally characterized into three subdivisions, and
they are namely: 1) fixed wing, 2) rotary wing, and 3) flapping wing.
The rotary wing UAVs (RUAVs) can be additionally categorized by the
number of rotors like a helicopter, quadcopter, hexacopter, octocopter,
and so forth.

Designing high-performance flight control systems for a UAV is a critical
and challenging task \cite{liu2015modeling}. There are some important
considerations in designing a reliable flight control system. The
first challenge is related to the robustness of the closed loop control
in the face of uncertainties, such as unpredictable external airflows
(e.g. severe wind gusts) and modelling error. The motion of a small
UAV can be highly vulnerable to the adverse impacts of wind gusts
that can force the system to depart from its desired trajectory. This
phenomena can lead to significant overshoots and tracking offsets,
which are undesirable in light of safety and efficiency issues. Significant
variations in plant dynamics (e.g. due to payload changes) can seriously
deteriorate the performance of fixed-gain control systems \cite{aastrom2013adaptive}.
To overcome this problem, a high-degree of flight autonomy is required
through the availability of robust and adaptive flight control systems.
It should be pointed out that one major problem with model-based control
systems is their dependency on the accuracy of the assumed mathematical
model of the system. In practice, there is no perfect mathematical
model to capture the whole dynamics of any systems, even for the simplest
ones. Although researchers have developed cutting-edge model-based
robust controls \cite{petersen2012robust}, the performance of linear
time-invariant (LTI) robust controls (e.g. H infinity and mu-synthesis\cite{santoso2015robust})
can deteriorate in the face of large uncertainties such as the failure
or substantial degradation of servos, control surfaces and sensors.

In such circumstances, approaches without the necessity of accurate
mathematical models of the system under control, are much appreciated.
Being a model-free approach, the Neural Network (NN) and Fuzzy Logic
system (FLS) based controllers have been successfully implemented
in many control applications \cite{pan2015peaking,pan2017biomimetic,pan2017composite}
over the past few years. To handle uncertainties in control system,
researchers have tried to combine the FLS, NN, FNN system with sliding
mode control (SMC), H$\infty$ control, back-stepping, etc. Such amalgamation
empowers the FLS, NN, FNN controller with the feature of tuning parameters,
which provides a more robust and adaptive control structure. However,
such adaptive FNN control structures are not able to evolve their
structures by adding or pruning rules. It forces the controller to
determine the number of rules a priori, where a selection of few fuzzy
rules may hinder to achieve adequate and desired control performance.
On the other hand, consideration of too many rules usually create
complex structures, which make them impossible to employ in real time.
A solution to the problem is utilization of evolving structure through
the addition or deletion of rules.

In recent time, researchers are trying to develop evolving FLS, NN,
FNN controllers by employing various approaches to add or delete the
rules \cite{pratama2014panfis,han2013real,han2014nonlinear,han2016nonlinear,han2015direct,dovvzan2014towards}.
In these controllers, the consequents are adapted using gradient-based
algorithms, evolutionary algorithms, or SMC theory. In case of gradient-based
algorithms slow convergence speed may be witnessed, in evolutionary
algorithm based controllers the stability proof is difficult and questionable
\cite{topalov2001online}, and SMC theory based adaptation methods,
there exists a dependency on PID parameters \cite{kayacan2013adaptive}.
To overcome these limitations, a new evolving controller namely Generic-controller
(G-controller) is proposed in our work, where the evolving architecture
is developed using an incremental learning method called GENEFIS \cite{pratama2014genefis},
and the consequents are adapted using SMC theory without any dependency
on the PID gains. Furthermore, the integration of GART+ in evolving
rules triggers a quick response, and a reduction in computational complexity
due to the unnecessary pruned rules. Utilization of self-organizing
sliding parameters is a newly used concept and contribution of our
work too.

The organization of the remaining part of the paper is as follows:
In Section \ref{sec:Problem-Formulation-in}, the formulation of the
hexacopter air vehicle, existing challenges in its control methods
are discussed. Section \ref{sec:Self-organizing-Mechanism-of} describes
the self-organizing mechanism of the G-controller. In the next section
\ref{sec:SMC-Theory-based-Adaptation}, the SMC theory based adaptation
of the consequent parameters are explained. The performance of the
proposed controller is summarized, compared and discussed in section
\ref{sec:Results-and-Discussion}. At last, the paper ends with the
concluding remarks mentioned in Section \ref{sec:Conclusion}.

\section{Problem Formulation in Unmanned Hexacopter Air Vehicle\label{sec:Problem-Formulation-in}}

\begin{figure*}[tbh]
\begin{centering}
\includegraphics[scale=0.33]{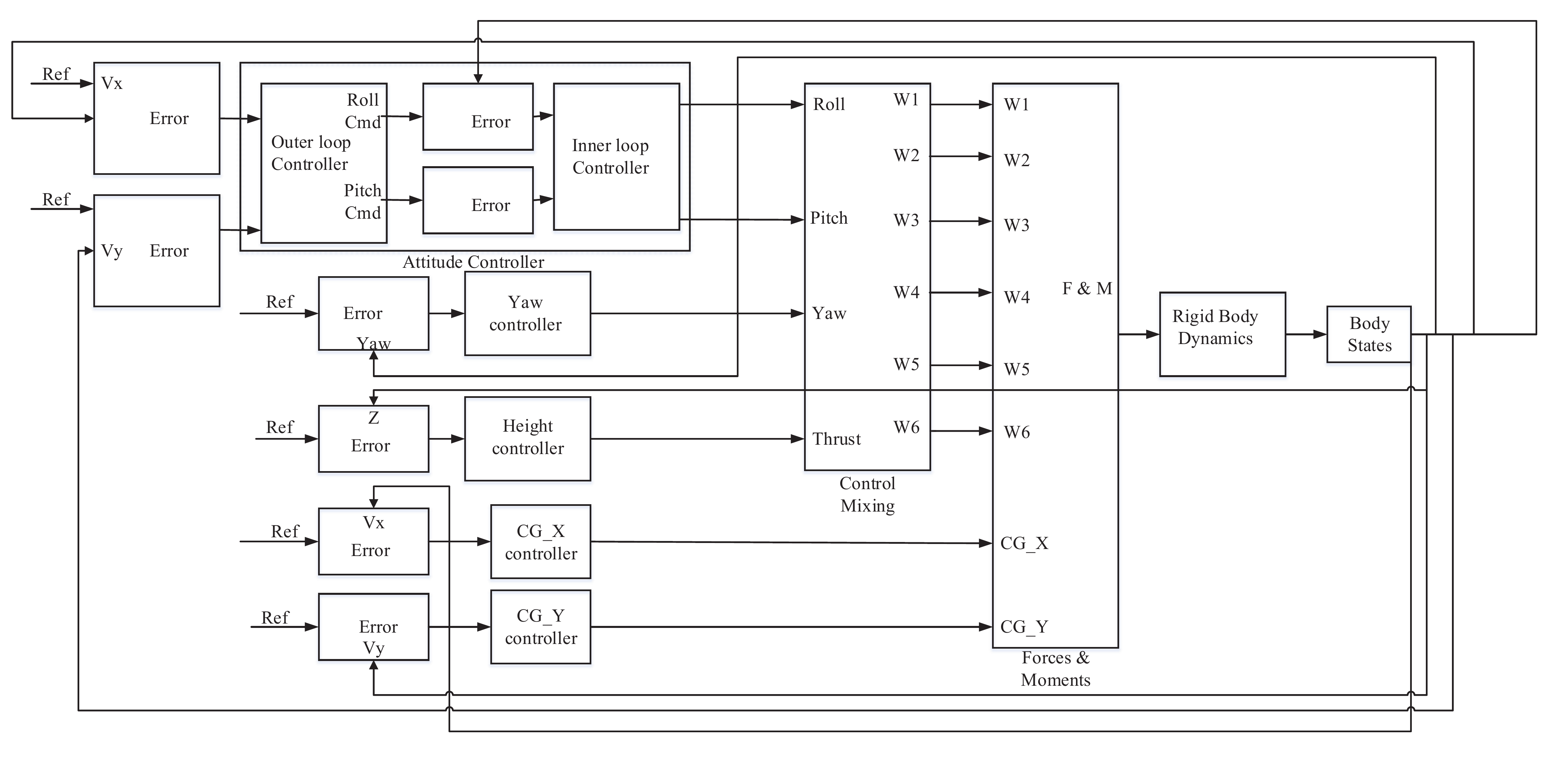}
\par\end{centering}

\caption{Top-level diagram of the over-actuated simulated Hexacopter plant
(adapted from \cite{ferdaus2018generic} with proper permission)\label{fig:Top-level-diagram-of}}

\end{figure*}

The simulated hexacopter plant is developed by UAV laboratory of the
UNSW at the Australian Defence Force Academy. The model is of medium
fidelity and contains both full 6 degrees of freedom (DOF) rigid body
dynamics and non-linear aerodynamics. The hexacopter simulated plant
introduces two extra degrees of freedom which are obtained by shifting
two masses using two aircraft servos with each mass sliding along
its own rail aligned in longitudinal and lateral directions respectively,
which makes the plant an over-actuated system. The top-level diagram
of this over-actuated simulated plant is exhibited in Fig. \ref{fig:Top-level-diagram-of}.
The mathematical modelling of the hexcopter along with problem formulation
is described in the following paragraphs of this section.

The body axes system of the hexacopter Unmanned Aerial Vehicle (UAV)
are fixed to the aircraft center of gravity and rotates as the aircraft\textquoteright{}s
attitude changes. This set of axes is particularly useful as the sensors
are fixed with respect to the body axes. Another set of axes, known
as the inertial axes, are required for navigation and defined with
respect to the surface of the earth. The inertial axes are aligned
so that the x-axis is horizontal and points North, the y-axis is horizontal
and points East and the z-axis is positive down towards the center
of the earth. A precise mapping between the inertial and body axes
can be made based on the attitude of the hexacopter.

Using standard aircraft nomenclature, the velocity components of the
hexacopter along the body axes x, y and z are given the designations
$u$, $v$ and $w$ respectively. Likewise, the body axes rotation
rates of the hexacopter are $p$, $q$ and $r$. The sense of the
rotations are defined in accordance with a right hand axes system.
In order to properly define the orientation of an aircraft it is not
only necessary to define a coordinate system about which to apply
rotations, but also the order in which they are applied. In aviation
three Euler angles are used to describe the orientation of an aircraft
with respect to an axes system fixed to the earth. These angles use
the familiar designations roll($\phi$), pitch($\theta$) and yaw($\psi$).
In order to avoid the wraparound problem and to linearise the attitude
update, the hexacopter model makes use of the four quaternion parameters
$q_{i}$ (where $i=0,1,2,3$) to store attitude and are converted
to the Euler angles as required using Eq. \ref{eq:q0_q3}.

\begin{equation}
\begin{array}{ccc}
q_{0} & = & \cos\frac{\phi}{2}\cos\frac{\theta}{2}\cos\frac{\psi}{2}+\sin\frac{\phi}{2}\sin\frac{\theta}{2}\sin\frac{\psi}{2}\\
q_{1} & = & \sin\frac{\phi}{2}\cos\frac{\theta}{2}\cos\frac{\psi}{2}-\cos\frac{\phi}{2}\sin\frac{\theta}{2}\sin\frac{\psi}{2}\\
q_{2} & = & \cos\frac{\phi}{2}\sin\frac{\theta}{2}\cos\frac{\psi}{2}+\sin\frac{\phi}{2}\cos\frac{\theta}{2}\sin\frac{\psi}{2}\\
q_{3} & = & \cos\frac{\phi}{2}\cos\frac{\theta}{2}\sin\frac{\psi}{2}-\sin\frac{\phi}{2}\sin\frac{\theta}{2}\cos\frac{\psi}{2}
\end{array}\label{eq:q0_q3}
\end{equation}

In general flight, rotors of the hexacopter experience a relative
freestream velocity due to its own motion of $V_{\infty}$. The airstream
is deflected through the actuator disc by speed $V_{i}$ at the disc
and can be shown to change the downstream flow by $2V_{i}$. This
flow is made up of components $V_{n}$ and $V_{t}$ perpendicular
and tangential to each rotor disk respectively. The values of $V_{n}$
and $V_{t}$ are calculated by adding the perpendicular and tangential
components of $V_{\infty}$ to the airflow created by pitching, rolling
and yawing motions at each rotor. In this modelling it is assumed
that the inflow $V_{i}$ does not change with radius or azimuth. The
elemental forces are integrated to achieve a closed form solution
for thrust in terms of blade pitch ($\theta_{0}$), inflow relative
to the rotor disk ($\lambda^{'}$) and advance ratio ($\mu$) as per
Eq. \ref{eq:T} as derived in \cite{seddon1990basic}.

\begin{equation}
T=\frac{\rho a(\Omega R)^{2}A_{b}}{2}\left[\frac{1}{3}\theta_{0}\left(1+\frac{3}{2}\mu^{2}\right)-\frac{1}{2}\lambda^{'}\right]\label{eq:T}
\end{equation}

where $\Omega$ is the blade rotational speed and

\begin{equation}
\lambda^{'}=\frac{V_{i}+V_{n}}{\Omega R}\;\;\mbox{and}\;\;\mu=\frac{V_{t}}{\Omega R}
\end{equation}

Based on the downwash for the equivalent wing, the mean induced velocity
$V_{i}$ is expressed as follows:

\begin{equation}
V_{i}=\frac{T}{2\rho A\hat{V}}\;\;\mbox{where}\;\;\hat{V}=\sqrt{V_{T}^{2}+(V_{n}+V_{i})^{2}}
\end{equation}

To obtain a good match between our developed hexacopter model and
experiments, Eq. 4 is modified in our work as follows:

\begin{equation}
V_{i}^{2}=\sqrt{\left(\frac{\hat{V}}{2}\right)^{2}+\left(\frac{T}{2\rho A}\right)^{2}}-\frac{\hat{V}^{2}}{2}
\end{equation}

The yawing torque $N$ generated by each rotor of the hexacopter results
from the drag of the rotor blades through the air. The rotor torque
is calculated by dividing the main rotor power $P_{tot}$ by the angular
velocity as follows

\begin{equation}
N=\frac{P_{tot}}{\Omega}
\end{equation}

The power $P_{tot}$ is due to a number of sources: the induced power
(Pind) which is the power required to create the induced velocity
$V_{i}$ and climb against gravity. And the profile power ($P_{0}$)
which is the power to overcome the profile drag of the blades. Therefore,
the power $P_{tot}$ is written compactly as follows

\begin{equation}
P_{tot}=P_{ind}+P_{0}
\end{equation}

where

\begin{equation}
P_{ind}=k_{ind}TV_{i}+TV_{c}
\end{equation}

\begin{equation}
P_{0}=\frac{\sigma C_{D_{0}}}{8}(1+\kappa\mu^{2})
\end{equation}

where $k_{ind}$ is a correction factor to compensate for non-uniform
induced velocity, tip loss effects etc. Likewise the constant $\kappa$
corrects for skewed flow and other effects in forward flight. $V_{c}$
represents the climb speed of the rotor. For the purposes of this
analysis we ignored the effects of vortex ring state in rapid descent.

The body of the hexacopter UAV is assumed to act as a rigid body.
Newton\textquoteright{}s second law of motion can be used to derive
the relationships between the forces and moments acting on the helicopter
and the linear and angular accelerations. Assuming that the hexacopter
is of a conventional mass distribution, it is usual that the xz plane
is a plane of symmetry, so that the cross product moments of inertia
$Iyz=Ixy=0$. In this case the equations of motion are those in Eq.
\ref{eq:F_M}. A good derivation of these equations are adopted from
the flight mechanics text by Nelson \cite{nelson1998flight}.

\begin{equation}
\begin{array}{ccc}
F_{x} & = & m(\dot{u}+qw\text{\textminus}rv)\\
F_{y} & = & m(\dot{v}+ru\text{\textminus}pw)\\
F_{z} & = & m(\dot{w}+pv\text{\textminus}qu)\\
L & = & I_{x}\dot{p}\text{\textminus}I_{xz}\dot{r}+qr(I_{z}\text{\textminus}I_{y})\text{\textminus}I_{xz}pq\\
M & = & I_{y}\dot{q}+rp(I_{x}\text{\textminus}I_{z})+I_{xz}(p^{2}\text{\textminus}r^{2})\\
N & = & \text{\textminus}I_{xz}\dot{p}+I_{z}\dot{r}+pq(I_{y}\text{\textminus}I_{x})+I_{xz}qr
\end{array}\label{eq:F_M}
\end{equation}

where

\begin{equation}
\begin{array}{ccc}
I_{x} & = & \int\int\int(y^{2}+z^{2})dm\\
I_{y} & = & \int\int\int(x^{2}+z^{2})dm\\
I_{z} & = & \int\int\int(x^{2}+y^{2})dm\\
I_{xy} & = & \int\int\int xydm\\
I_{xz} & = & \int\int\int xzdm\\
I_{yz} & = & \int\int\int yzdm
\end{array}
\end{equation}

The mass $m$ and mass moments of inertia $Ix,\: Iy,\: Iz$ and $I_{xz}$
of the hexacopter are given in table \ref{tab:Hexacopter-inertia-properties}.

\begin{center}
\begin{table}[tbh]
\caption{Hexacopter inertia properties\label{tab:Hexacopter-inertia-properties}}

\centering{}%
\begin{tabular}{|c|c|c|c|}
\hline
\textbf{Parameter} & \textbf{Description} & \textbf{Units} & \textbf{Value}\tabularnewline
\hline
$m$ & mass & $kg$ & 3.0\tabularnewline
\hline
$I_{x}$ & Mass Moment about x-axis & $kgm^{2}$ & 0.04\tabularnewline
\hline
$I_{y}$ & Mass Moment about y-axis & $kgm^{2}$ & 0.04\tabularnewline
\hline
$I_{z}$ & Mass Moment about z-axis & $kgm^{2}$ & 0.06\tabularnewline
\hline
$I_{xz}$ & Product of Inertia & $kgm^{2}$ & 0\tabularnewline
\hline
$g$ & Gravitational constant & $ms^{-2}$ & 9.81\tabularnewline
\hline
\end{tabular}
\end{table}

\par\end{center}

For robustness, the attitude of the hexacopter is stored as a quaternion
and updated using the following Eq. \ref{eq:q0_q4_v2} provided in
\cite{stevens2015aircraft}. The quaternion attitude update also removes
the need to use trigonometric functions which would be required if
integrating the Euler angle differential equations.

\begin{equation}
\left[\begin{array}{c}
q_{0}\\
q_{1}\\
q_{2}\\
q_{3}
\end{array}\right]=-\frac{1}{2}\left[\begin{array}{cccc}
0 & p & q & r\\
-p & 0 & -r & q\\
-q & r & 0 & -p\\
-r & -q & p & 0
\end{array}\right]\label{eq:q0_q4_v2}
\end{equation}

The final step in updating the rigid body states is to update the
position of the hexacopter in global coordinates relative to an earth-based
axes system. The local velocities $u$, $v$ and $w$ are first converted
to global velocities $\dot{X}$ , $\dot{Y}$ and $\dot{Z}$ by multiplying
the local velocities by the rotation matrix $B$ as in Eq. \ref{eq:vel_3d}.
The rotation matrix can be determined directly from the quaternions
using Eq. \ref{eq:B}. These velocities are then integrated to obtain
the global position $[X,\: Y,\: Z]$.

\begin{equation}
\begin{array}{ccc}
\left[\begin{array}{c}
\dot{X}\\
\dot{Y}\\
\dot{Z}
\end{array}\right] & = & B\left[\begin{array}{c}
u\\
v\\
w
\end{array}\right]\end{array}\label{eq:vel_3d}
\end{equation}

where

\begin{equation}
\begin{array}{lcc}
B=\\
\left[\begin{array}{ccc}
q_{0}^{2}+q_{1}^{2}\text{-}q_{2}^{2}-q_{3}^{2} & 2(q_{1}q_{2}+q_{0}q_{3}) & 2(q_{1}q_{2}+q_{0}q_{3})\\
2(q_{1}q_{2}\text{\textminus}q_{0}q_{3}) & q_{0}^{2}+q_{1}^{2}\text{-}q_{2}^{2}\text{-}q_{3}^{2} & 2(q_{2}q_{3}+q_{0}q_{1})\\
2(q_{1}q_{3}+q_{1}q_{2}) & 2(q_{2}q_{3}\text{\textminus}q_{0}q_{1}) & q_{0}^{2}-q_{1}^{2}\text{\textminus}q_{2}^{2}+q_{3}^{2}
\end{array}\right]
\end{array}\label{eq:B}
\end{equation}

Eq. \ref{eq:F_M} have been implemented as a C code SIMULINK R S-function
in the simulated hexacopter plant. The states for the dynamics block
are position, local velocity components in the hexacopter axes system,
rotation rates and quaternion attitude. Inputs to the block are the
forces and moments acting on the hexacopter while the outputs are
accelerations, local velocities, position, body angular rates and
attitude. From the body states block, the body position of the hexacopter
in Z- axes i.e. the altitude $Z$ and attitude (roll and pitch) is
supplied to the error calculation block. From here the error is calculated
by measuring the difference between actually obtained altitude, attitude
and desired reference altitude, attitude. This error is supplied to
the proposed G-controller. To prove the closed-loop stability by considering
both plants and controller, SMC theory based sliding surface is utilized
in this work, where there is no plant parameter dependency. It makes
our proposed G-controller a plant parameter free i.e. a real model-free
controller.

\section{Self-organizing Mechanism of G-controller\label{sec:Self-organizing-Mechanism-of}}

The G-controller is built using a self-evolving fuzzy system namely
GENEFIS \cite{pratama2014genefis}, where GENEFIS is an evolving TS fuzzy system with
ellipsoidal contours in arbitrary positions. a typical fuzzy rule
of the G-controller can be presented as follows:

\begin{equation}
\text{IF}~Z~\text{is}~R_{i},~\text{then}~y_{i}=b_{0i}+b_{1i}\zeta_{1}+b_{2i}\zeta_{2}+...+b_{ki}\zeta_{k}\label{eq:3}
\end{equation}
where $R_{i}$ denotes the $i-th$ rule (membership function) constructed
from a concatenation of fuzzy sets and epitomizing a multidimensional
kernel, $k$ represents the dimension of input feature, $Z$ is an
input vector of interest, $b_{i}$ is the consequent parameter, $\zeta_{k}$
is the $k-th$ input feature. The predicted output of the self-evolving
model can be expressed as:

\begin{align}
\hat{y} & =\sum\limits _{i=1}^{j}\psi_{i}(\zeta)y_{i}(\zeta)=\frac{\sum\limits _{i=1}^{j}R_{i}y_{i}}{\sum\limits _{i=1}^{j}R_{i}}\nonumber \\
 & =\frac{\sum_{i=1}^{j}\text{exp}(-(Z-\Theta_{i})\Sigma_{i}^{-1}(Z-\Theta_{i})^{T})y_{i}}{\sum_{i=1}^{j}\text{exp}(-(Z-\Theta_{i})\Sigma_{i}^{-1}(Z-\Theta_{i})^{T})}\label{eq:y1}
\end{align}
In Eq. \ref{eq:y1}, $\Theta_{i}$ is the centroid of the $i-th$
fuzzy rule $\Theta_{i}\in\Re^{1\times j}$, $\Sigma_{i}$ is a non-diagonal
covariance matrix $\Sigma_{i}\in\Re^{k\times k}$ whose diagonal components
are expressing the spread of the multivariate Gaussian function, and
$k$ is the number of fuzzy rules.

\subsection{Mechanism of Online Rule-Growing }

The Datum Significance (DS) method developed in \cite{huang2004efficient}
is extended in \cite{pratama2014panfis} to cope with multivariate
Gaussian membership function, which is utilized in our work as a rule-growing
mechanism. In that regards, after several mathematical amendment to
the original DS method, the expression is as follows:

\begin{equation}
D_{sig}=|e_{rn}|\frac{\text{det}(\Sigma_{j+1})^{k}}{\sum_{i=1}^{j+1}\text{det}(\Sigma_{i})^{k}}\label{eq:DS2}
\end{equation}

When a outlier is obtained far away from the nearest rule, a high
value of $D_{sig}$ may obtain from Eq. $\ref{eq:DS2}$ even with
a small value of $e_{rn}$. Besides, the obtained $e_{rn}$ may have
a high value in overfitting situation. In such situation, a newly-added
rule worsen the situation. As a solution to the above mentioned problems,
Eq. $\ref{eq:DS2}$ needs to be separated.

In our work the rule growing method is activated when the rate of
change of $e_{rn}$ is positive, where mean and variance of $e_{rn}$
is updated recursively \cite{angelov2011fuzzily} as follows:

\begin{equation}
\bar{e}_{rn}=\frac{n-1}{n}\bar{e}_{rn-1}+\frac{1}{k}\bar{e}_{rn}
\end{equation}

\begin{equation}
\bar{\sigma}_{rn}^{2}=\frac{n-1}{n}\bar{\sigma}_{rn-1}^{2}+\frac{1}{k}(\bar{e}_{rn}-\bar{e}_{rn-1})
\end{equation}
When the condition $\bar{e}_{rn}+\bar{\sigma}_{rn}^{2}-(\bar{e}_{rn-1}+\bar{\sigma}_{rn-1}^{2})>0$
is fulfilled, the DS criterion presented in Eq. \ref{eq:DS2} is simplified
here as follows:

\begin{equation}
D_{sig}=\frac{\text{det}(\Sigma_{j+1})^{k}}{\sum_{i=1}^{j+1}\text{det}(\Sigma_{i})^{k}}\label{eq:DS3}
\end{equation}
When the $D_{sig}$ calculated in Eq. \ref{eq:DS3} satisfied the
condition $D_{sig}\ge g$, the rule base is expanded. Here $g$ is
a predefined threshold. The possibility of overfitting phenomenon
due to a new rule is omitted by using Eq. \ref{eq:DS3}. Besides,
this DS criterion can predict the probable contribution of the datum
during its lifetime.

\subsection{Mechanism of Pruning Rule}

Numerous modifications are made in Extended Rule significance (ERS)
theory to fit them with the proposed G-controller. By using the $k-$fold
numerical integration, the final expression of ERS theory utilized
in our work is as follows:
\begin{equation}
E_{inf}^{i}=\sum_{i=1}^{j+1}\eta_{i}\frac{\text{det}(\Sigma_{i})^{k}}{\sum_{i=1}^{j}\text{det}(\Sigma_{i})^{k}}\label{eq:E_3}
\end{equation}
When $E_{inf}^{i}\le k_{e}$ i.e. the volume of the $i$th cluster
is much lower than the summation of volumes of all cluster, the rule
is considered as inconsequential and pruned to protect the rule base
evolution from its adverse effect. In this work, $k_{e}$ exhibits
a plausible trade-off between compactness and generalization of the
rule base. The allocated value for $\delta$ is $\delta=[0.0001,1]$,
and $k_{e}=10\%\;\text{of}\;\delta$.

\subsection{Adaptation of Rule Premise Parameters}

Generalized Adaptive Resonance Theory+ (GART+) \cite{oentaryo2011bayesian}
is used in G-controller as a technique to adapt premise parameters.
To relieve from the $\textit{cluster delamination}$ effect, in GENEFIS
based G-controller the size of fuzzy rule are constrained by using
GART+, which allows a limited grow or shrink of a category. The procedure
and conditions of selecting the winning rule using GART+ has explained
briefly in our previous work \cite{ferdaus2018generic} and in \cite{pratama2014genefis}.

\section{SMC Theory-based Adaptation in G-Controller\label{sec:SMC-Theory-based-Adaptation}}

In our proposed G-controller, the sliding mode control (SMC) theory
is applied to adapt the consequent parameters, which can guarantee
the robustness of a system against external perturbations, parameter
variations, and unknown uncertainties. Therefore, the SMC theory-based
adaptation laws are derived to develop the closed-loop system stability.
The zero dynamics of the learning error coordinate \cite{yu2009sliding,kayacan2017type}
is defined as time-varying sliding surface as follows:

\begin{equation}
S_{ssr}(u_{g},u)=u_{ARC}(t)=u_{g}(t)+u(t)
\end{equation}

The sliding surface for an over-actuated unmanned aerial vehicle namely
hexacopter to be controlled is expressed as:

\begin{equation}
s_{H}=e+\lambda_{1}\dot{e}+\lambda_{2}\int_{0}^{t}e(\tau)d\tau
\end{equation}
where, $\lambda_{1}=\frac{\alpha_{2}}{\alpha_{1}},$ $\lambda_{2}=\frac{\alpha_{3}}{\alpha_{1}}$$,$$\: e$
is the error which is the difference between the actual displacement
from the hexacopter plant and desired height. In this work, the sliding
parameter $\alpha_{1}$ has initialized with a small value $1\times10^{-6}$,
whereas $\alpha_{2}$ has initialized with $1\times10^{-6},$ and
$\alpha_{3}\thickapprox0.$ Each of the parameters is then evolved
by using different learning rates. These learning rates are set in
such a way so that the sliding parameters can achieve the desired
value in the shortest possible time to create a stable closed-loop
control system. A higher initial value of the sliding parameters is
avoided, since it may cause a big overshoot at the beginning of the
trajectory. In short, to make our proposed G-controller absolutely
model free, these sliding parameters are self-organizing rather than
predefined constant values.

The adaptation laws for the consequent parameters of the G-controller
are chosen as:

\begin{equation}
\dot{\omega}(t)=-\alpha_{1}G(t)\psi(t)s_{H}(t),\;\;\text{where }\;\omega(0)=\omega_{0}\in\Re^{nR\times1}
\end{equation}
where the term $G(t)$ can be updated recursively as follows:

\begin{equation}
\dot{G}(t)=-G(t)\psi(t)\psi^{T}(t)G(t),\;\;\text{where }\; G(0)=G_{0}\in\Re^{nR\times nR}
\end{equation}
where $n$ is the number of inputs to the controller, and $R$ is
the number of generated rules. These adaptation laws guarantee a stable
closed-loop control system, and explained with the stability proof
in \cite{ferdaus2018generic}.

\section{Results and Discussion\label{sec:Results-and-Discussion}}

\begin{figure}[tbh]
	\begin{centering}
		\includegraphics[scale=0.17]{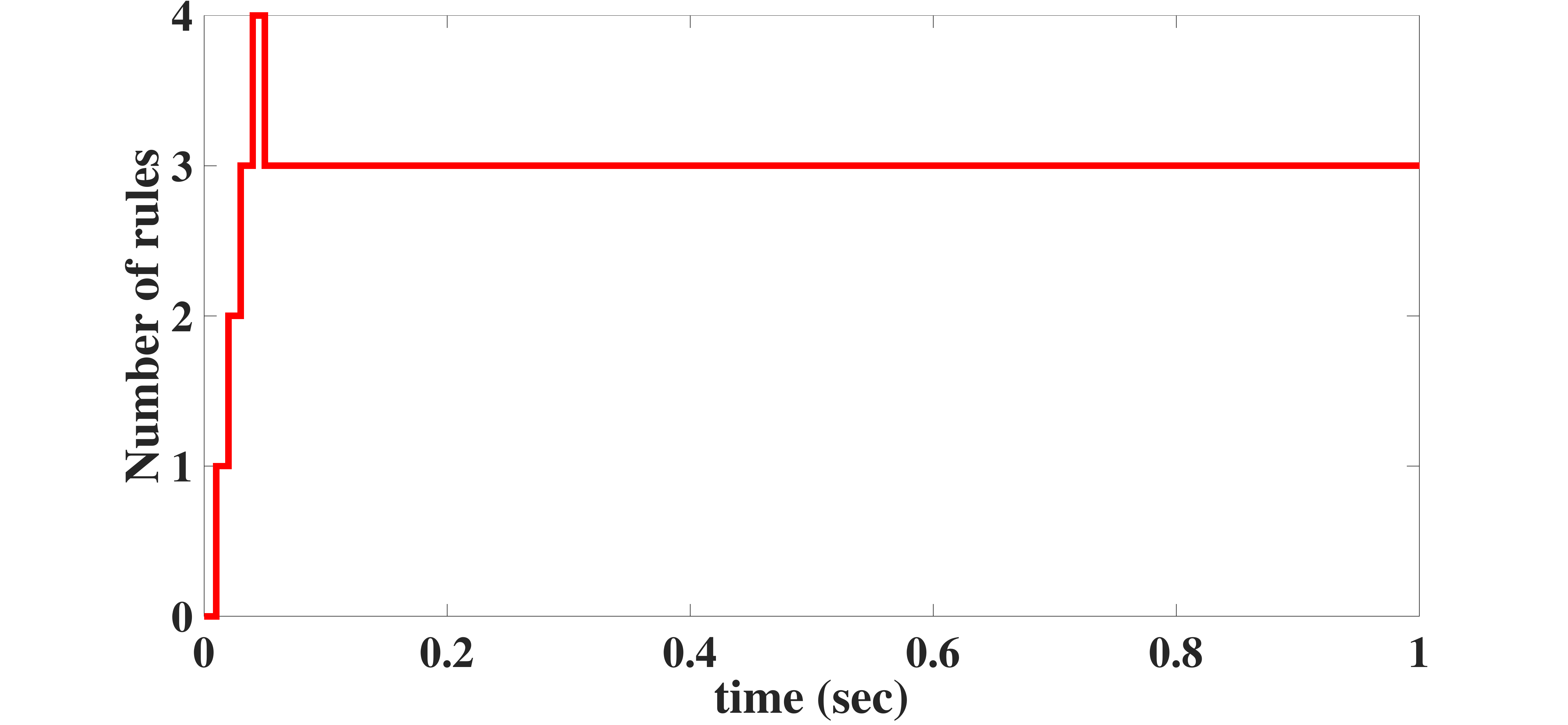}
		\par\end{centering}
	
	\caption{Rule evolving phenomenon of the self-evolving G-controller in case of the constant
		altitude trajectory of the hexacopter\label{fig:Rule-evolution-of}}

\end{figure}

In our work, the self-evolving generic neuro-fuzzy controller namely
G-controller is utilized to control a six-rotored UAV namely hexacopter.
A variety of altitude trajectory tracking is witnessed to evaluate
the controller's performance. Being an self-organizing evolving controller,
the G-controller can evolve both the structure and parameters by adding
or pruning the rules like many other evolving controllers discussed
in the section \ref{sec:Introduction}. To understand clearly, one of the rule evolving phenomenon in case of constant altitude is presented graphically in Fig. \ref{fig:Rule-evolution-of}. Nonetheless, the addition
of GRAT+, multivariate Gaussian function, SMC learning theory based
adaptation laws support it to provide an improved trajectory tracking
performance. The controllers are employed to control the thrust of
the control-mixing box of the hexacopter plant.

\begin{figure}[tbh]
\begin{centering}
\subfloat[]{\includegraphics[scale=0.17]{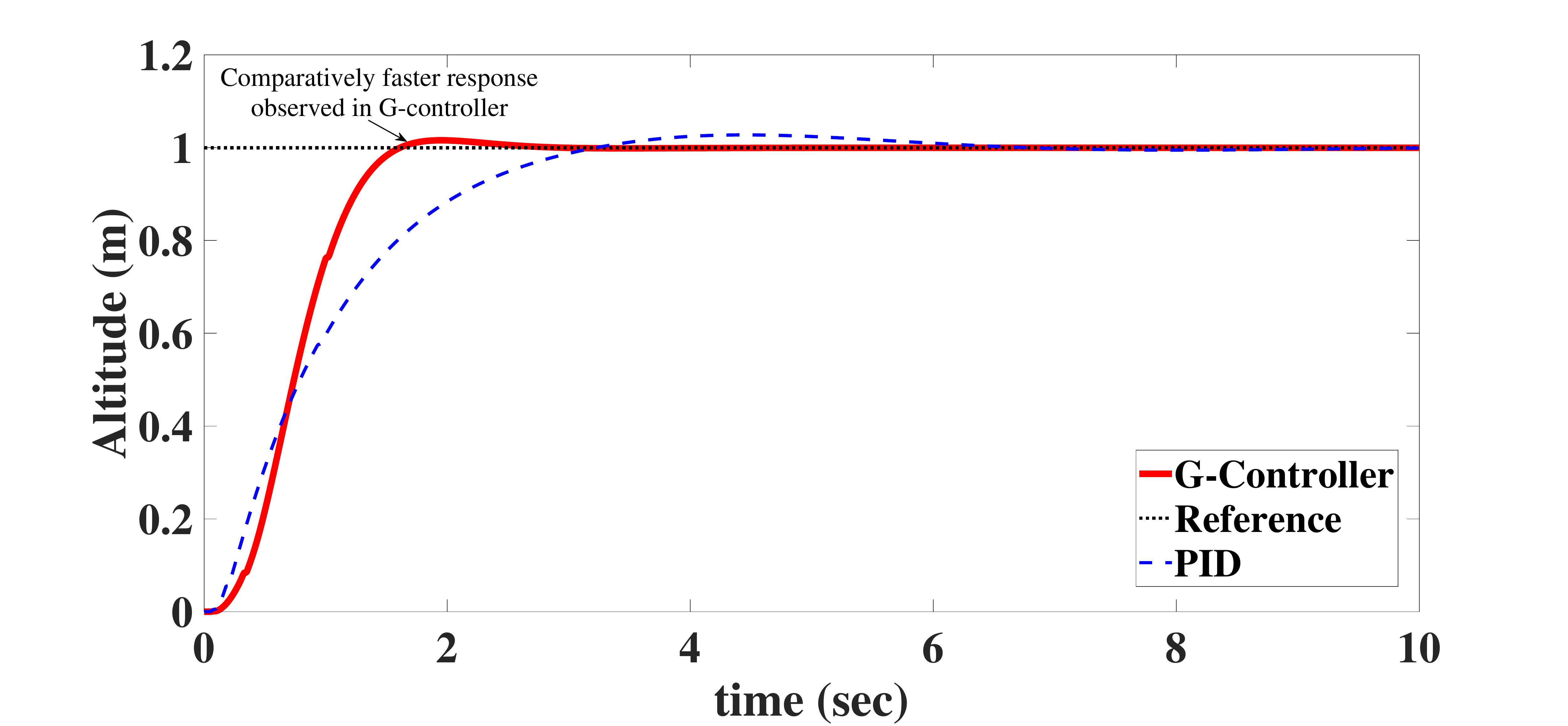}

}
\par\end{centering}

\begin{centering}
\subfloat[]{\includegraphics[scale=0.17]{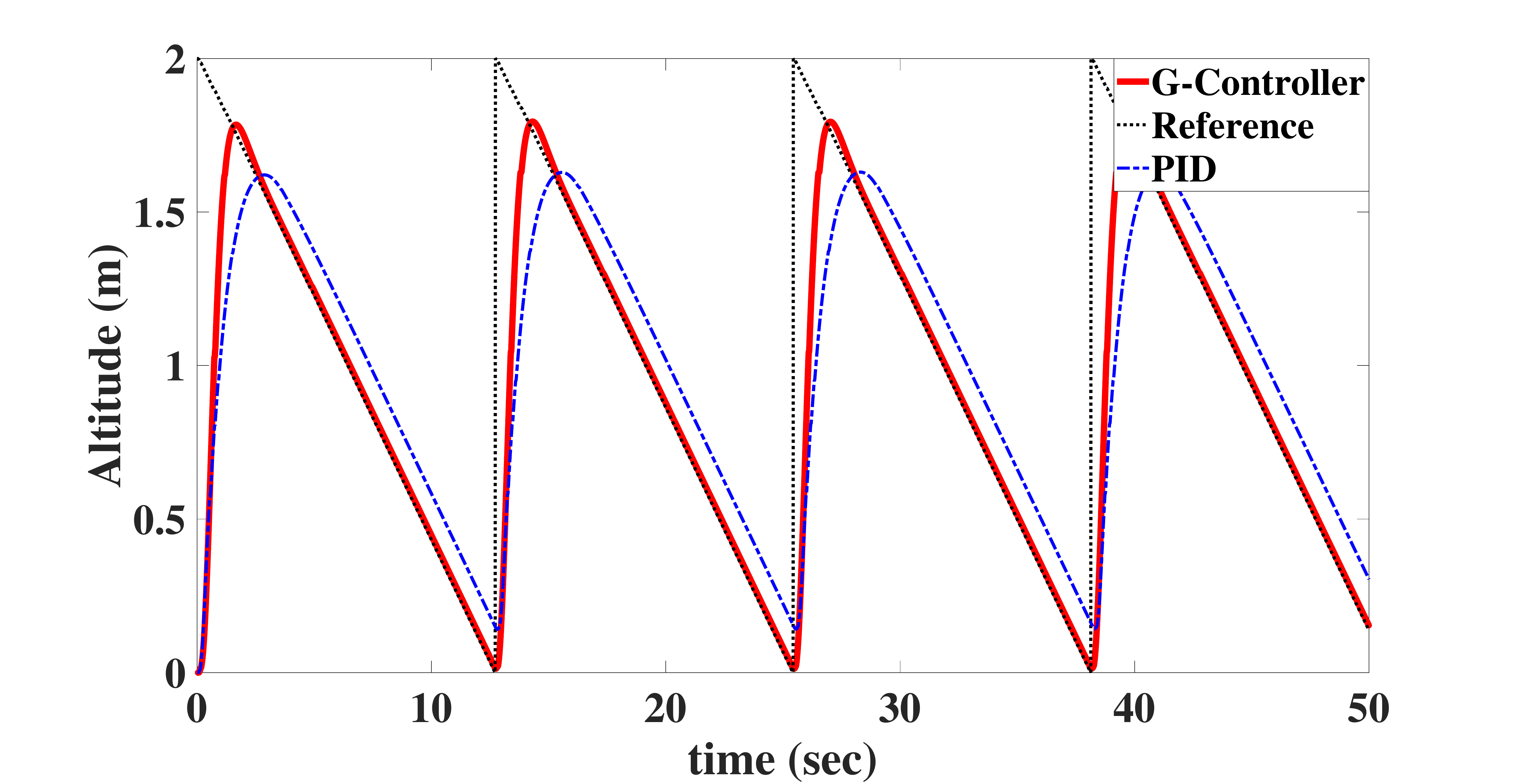}

}
\par\end{centering}

\begin{centering}
\subfloat[]{\includegraphics[scale=0.17]{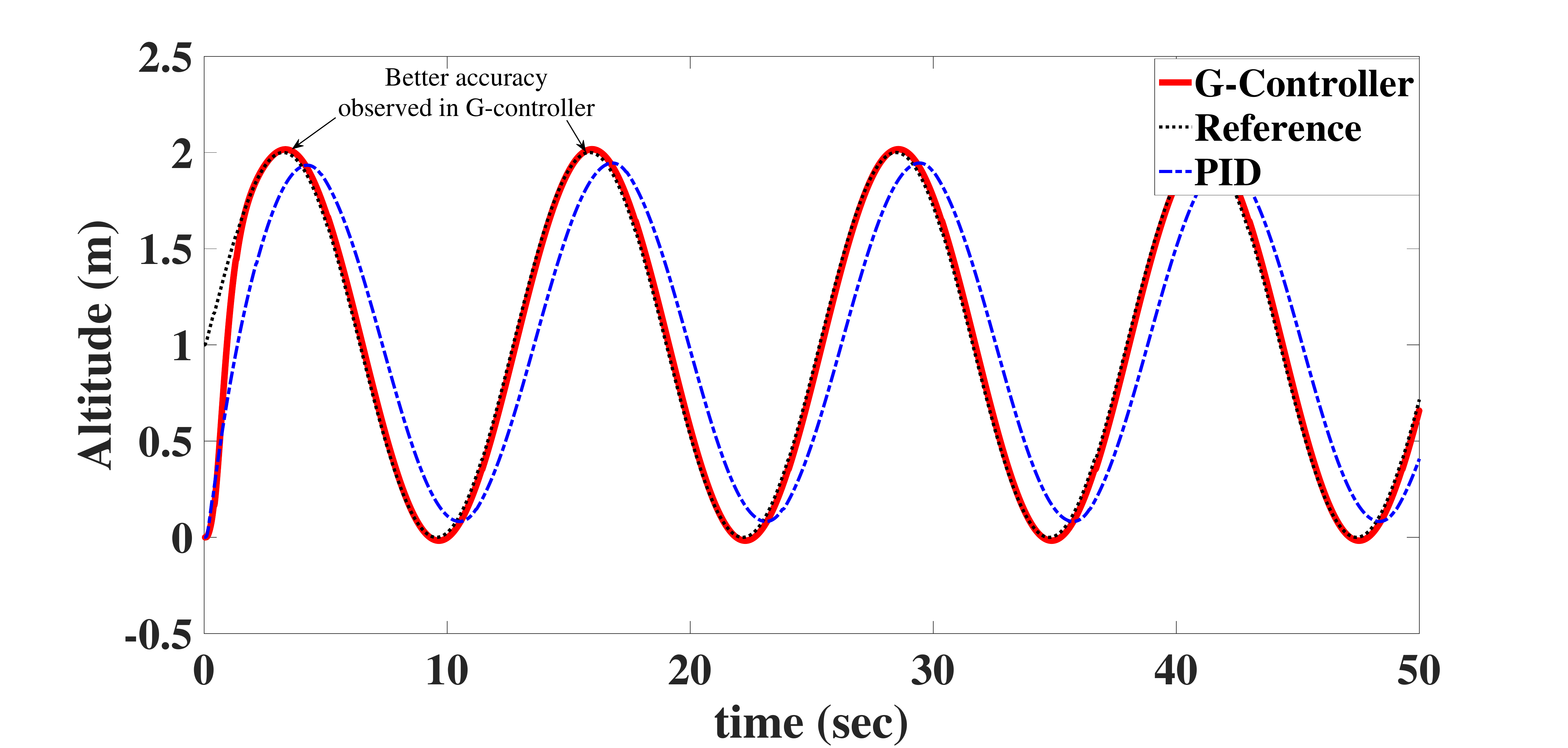}

}
\par\end{centering}

\begin{centering}
\subfloat[]{\includegraphics[scale=0.17]{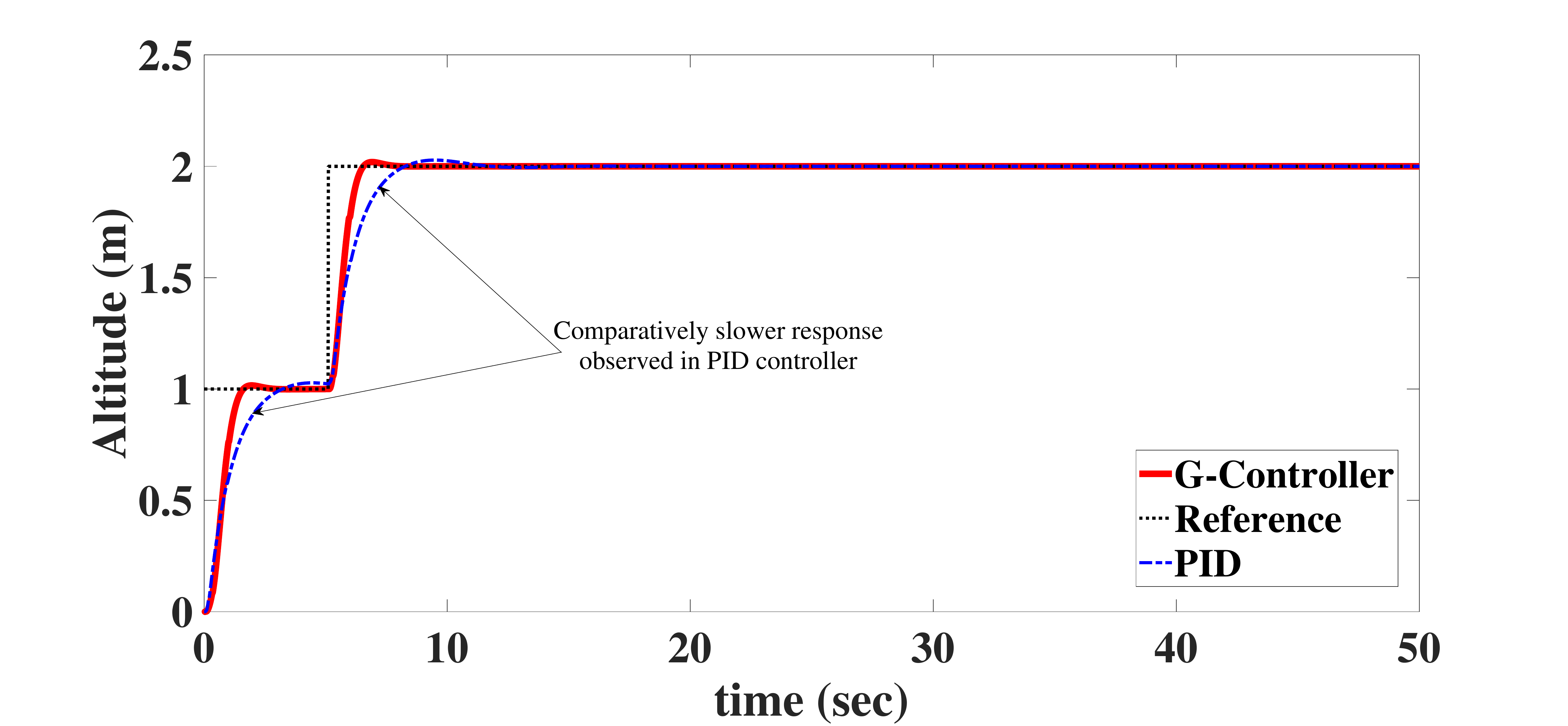}

}
\par\end{centering}

\caption{Performance observation of a PID and proposed G-controller in tracking
various altitude of the hexacopter\label{fig:Performance-observation-of_altitude}}
\end{figure}

The G-controller's performance is observed with respect to various
reference altitude such as: 1) a constant altitude of 1 meter expressed
as $Z_{d}(t)=1$; 2) a triangle wave function with a frequency of
0.1 Hz, and amplitude of 2 m; 3) a sine wave function with a frequency
of 0.1 Hz, and amplitude of 2 m; and 4) a step function presented
as $Z_{d}(t)=u(t)+u(t-5)$. All these results are compared with a
Proportional Integral Derivative (PID) controller. The altitude tracking
performance of the proposed controller for various trajectories has
been observed in Fig. \ref{fig:Performance-observation-of_altitude}.
In all cases, better tracking has been observed from the G-controller
than the PID controller. The RMSE, rise time, and settling time has
been calculated for all these trajectories and outlined in TABLE \ref{tab:hexa_RMSE}.
The lower RMSE is obtained from the proposed G-controller. Besides,
the settling time of the G-controller is much lower than the PID,
which clearly indicates its improvement over the PID controller.

\begin{center}
\begin{table}[t]
\caption{Measured RMSE, rising and settling time of various controllers in
controlling various altitude of the hexacopter\label{tab:hexa_RMSE}}

\centering{}%
\begin{tabular}{|>{\raggedright}p{0.85cm}|>{\raggedright}p{0.99cm}|>{\raggedright}p{1.1cm}|>{\raggedright}m{1.1cm}|>{\raggedright}p{0.7cm}|>{\raggedright}p{1cm}|}
\hline
\multirow{4}{0.85cm}{\textbf{Hexa-copter movement}} & \multirow{4}{0.99cm}{\textbf{Desired trajectory}} & \multirow{4}{1.1cm}{\textbf{Maximum amplitude \& angle}} & \multirow{4}{1.1cm}{\textbf{Measured feature}} & \multirow{4}{0.7cm}{\textbf{PID}} & \multirow{4}{1cm}{\textbf{G-controller}}\tabularnewline
 &  &  &  &  & \tabularnewline
 &  &  &  &  & \tabularnewline
 &  &  &  &  & \tabularnewline
\hline
\multirow{12}{0.85cm}{\textbf{Altitude}} & \multirow{3}{0.99cm}{Constant amplitude} & \multirow{3}{1.1cm}{1 m} & \textbf{RMSE} & 0.2456 & \textbf{0.2417}\tabularnewline
\cline{4-6}
 &  &  & \textbf{Rise time (sec)} & 3.2291 & \textbf{1.6112}\tabularnewline
\cline{4-6}
 &  &  & \textbf{Settling time (sec)} & 7.5112 & \textbf{2.9145}\tabularnewline
\cline{2-6}
 & \multirow{3}{0.99cm}{Step function} & \multirow{3}{1.1cm}{2 m} & \textbf{RMSE} & 0.1535 & \textbf{0.1445}\tabularnewline
\cline{4-6}
 &  &  & \textbf{Rise time (sec)} & 3.2221 & \textbf{1.6111}\tabularnewline
\cline{4-6}
 &  &  & \textbf{Settling time (sec)} & 12.1211 & \textbf{3.0100}\tabularnewline
\cline{2-6}
 & \multirow{3}{0.99cm}{Sine wave function} & \multirow{3}{1.1cm}{2 m} & \textbf{RMSE} & 0.3335 & \textbf{0.1283}\tabularnewline
\cline{4-6}
 &  &  & \textbf{Rise time (sec)} & 4.0200 & \textbf{1.9410}\tabularnewline
\cline{4-6}
 &  &  & \textbf{Settling time (sec)} & 4.1000 & \textbf{2.0100}\tabularnewline
\cline{2-6}
 & \multirow{3}{0.99cm}{Sawtooth wave function} & \multirow{3}{1.1cm}{2 m} & \textbf{RMSE} & 0.4391 & \textbf{0.3930}\tabularnewline
\cline{4-6}
 &  &  & \textbf{Rise time (sec)} & 2.5710 & \textbf{1.5420}\tabularnewline
\cline{4-6}
 &  &  & \textbf{Settling time (sec)} & 3.6120 & \textbf{2.7100}\tabularnewline
\hline
\end{tabular}
\end{table}

\par\end{center}

\section{Conclusion\label{sec:Conclusion}}

Being self-organizing and evolving in nature, our proposed G-controller
can evolve both the structure and parameters. To increase this controller's
robustness against uncertainties SMC theory based adaptation laws
are synthesized too. These are some desirable characteristics to control
a highly nonlinear UAV like hexacopter. Therefore, in this work the
G-controller based high performance closed-loop altitude control system
is developed to track the desired trajectory. In this work, our proposed
control algorithm is developed using C programming language considering
the compatibility issues to implement directly in hardware of hexacopter and the code is made available online in \cite{mpratamaweb}.
The performances are compared with a PID controller and improved results
are observed from our proposed G-controller. The controller starts
building the structure from scratch with an empty fuzzy set in the
closed-loop system. It causes a slow response at the starting point
for a very insignificant time, which is a common phenomenon in any
self-evolving controller. Nevertheless, the amalgamation of GRAT+,
multivariate Gaussian function, SMC learning theory based adaptation
laws, the self-evolving mechanism in the G-controller make it faster
with a lower computational cost. In addition, the G-controller's stability
is confirmed by both the Lyapunov theory and experiments. In future,
the controller will be utilized through hardware-based flight test
of various unmanned aerial vehicles.

\section*{Acknowlegment}
This research is fully supported by NTU start-up grant and MOE Tier-1 grant.
\bibliographystyle{IEEEtran}
\bibliography{conf4ref}



\end{document}